\documentclass[prd,showpacs,preprintnumbers,floatfix,superscriptaddress,10pt]{revtex4}
\usepackage[mathscr]{eucal}
\usepackage{color}
\usepackage[dvips]{graphicx}
\usepackage{epsf}
\usepackage{bm}

\newcommand{\be}{\begin{equation}}
\newcommand{\ee}{\end{equation}}
\newcommand{\ba}{\begin{eqnarray}}
\newcommand{\ea}{\end{eqnarray}}

\begin{document}

\title{Bulk viscosity coefficients   due to phonons and kaons  in superfluid color-flavor locked quark matter}

\author{Robert Bierkandt}
\affiliation{ Department of Astronomy and Astrophysics\\
Berlin Institute of Technology\\
Hardenbergstr. 36\\
D-10623 Berlin, Germany}

\author{Cristina Manuel}

\affiliation{ Instituto de Ciencias del Espacio (IEEC/CSIC) \\
Campus Universitat Aut\` onoma de Barcelona,
Facultat de Ci\` encies, Torre C5 \\
E-08193 Bellaterra (Barcelona), Spain}

\begin{abstract}
We evaluate the three bulk viscosity coefficients $\zeta_1, \zeta_2$ and $\zeta_3$ in the color-flavor locked (CFL) superfluid phase due to 
phonons and kaons, which are the lightest modes in that system. We first show that the computation is rather analogous to the computation
of the same coefficients in superfluid $^4$He, as due to phonons and rotons. For astrophysical applications, we also find the value of the
viscosities when there is a periodic disturbance, and the viscosities also depend on the frequency of the disturbance.
 In a temperature regime that might be of astrophysical relevance,  we find that the contributions of both the phonons and kaons should be
 considered, and that $\zeta_2$ is much less that the same coefficient in unpaired quark matter.
 \end{abstract}
\date{14th June 2011}


\maketitle

\section{Introduction}

  The ground state of three-flavor quark matter  at asymptotically large densities and low temperature  is  the color-flavor locked (CFL) phase~\cite{Alford:1998mk}.
 Many properties of CFL quark matter have been studied up to now ~\cite{reviews}.  
 The transport or kinetic coefficients of CFL quark matter are of particular interest, as  they may allow us to identify whether the CFL phase is realized or not
 in the core of compact stars, even in the situation where the density is not so large.  In the literature one can find computations of the shear viscosity \cite{Manuel:2004iv}, bulk viscosities \cite{Alford:2007rw,Manuel:2007pz,Mannarelli:2009ia},  thermal conductivity
 \cite{Shovkovy:2002kv,Braby:2009dw}, mutual friction \cite{Mannarelli:2008je}, as well as a general discussion on the kinetics
 of a CFL superfluid \cite{Mannarelli:2008jq}. 
 
CFL quark matter is a superfluid. The hydrodynamical description of superfluidity differs from that corresponding to a normal fluid, as in that case
one has to describe the flow corresponding to the quantum condensate, and the flow associated to the thermal quasiparticles of the system \cite{landaufluids,IntroSupe}.
Due to the existence of two different flows, more equations than those corresponding to a normal fluid are needed.  
In the dissipative regime of the superfluid one can also define more transport coefficients than for a normal fluid.
 In particular, it is known that due to the existence of these two different flows one can define up to  three bulk viscosity coefficients in a superfluid system.

In this paper we study the three bulk viscosities of CFL quark matter due to superfluid phonons and kaons. These collective excitations are the
Goldstone modes associated with the spontaneous breaking of baryon and of  chiral symmetry, respectively. These are the lightest modes in the CFL
phase, and thus, one can expect that they provide the leading contribution to its transport properties. The contribution to one of the bulk viscosity
coefficients due to kaons only was computed in Ref.~\cite{Alford:2007rw}. The most relevant process for that contribution comes from the decay
of the neutral kaons into superfluid phonons.
 In Ref.\cite{Mannarelli:2009ia}  the phonon contribution to the three bulk viscosity coefficients
was evaluated in a temperature regime where the kaon population was assumed to be thermally suppressed. In that case, the most relevant process that contributes
to the viscosities is one that changes the phonon number density, and implies five phonon collisions.
Because the kaon masses are not well-known,
it is not clear at which temperatures one should consider the kaon contribution into account.

In this article we compute the  three bulk viscosity coefficients of superfluid CFL quark matter due to kaons and phonons, 
 assuming a value of the kaon mass gap
within a reasonable range of values.  While one could naively think that
the result is the sum of the contributions of the kaons and the phonons, (roughly speaking, adding up the results of Ref.~\cite{Alford:2007rw} and Ref.\cite{Mannarelli:2009ia})
 we will show that this is not the case, the reason being that kaon decay process also changes the phonon number density.
 We will first show that the computation of the bulk viscosities  is totally analogous to that
corresponding to the phonon and roton contribution to the bulk viscosities in superfluid $^4$He. Further, and having in mind possible astrophysical applications,
 we will compute the viscosities when there is a periodic disturbance, when the viscosities depend also on the frequency of the disturbance. Obtaining the
 algebraic and numerical values of the viscosities is  the final goal of this manuscript.

This paper is structured as follows. In Section \ref{hydro} we give a very brief review to the relativistic hydrodynamic equations, and how the bulk viscosities
can be interpreted as coefficients that parametrize the deviations of both the pressure and chemical potential from its equilibrium values. In Section ~\ref{decay-sec}
we also review the two different processes and associated decay rates which contribute the most to the computation of the bulk viscosities. These are a five phonon
process, and the (electroweak mediated) kaon decay into two superfluid phonons. In Section \ref{bulk-sec} we compute the three viscosity coefficients when there
is a periodic disturbance, as a function of the frequency of this. In Subsection \ref{algebraic} we present the analytical results, while the numerical results are displayed
in Subsection \ref{numerical}.  Finally, we summarize our main findings in Section~\ref{conclu}. We will use natural units $\hbar = c  = k_B= 1$ everywhere in the manuscript, except when we show the numerical values of the viscosities.

\section{ Bulk viscosity coefficients in a relativistic superfluid}
\label{hydro}

The hydrodynamical equations governing the bulk fluctuations of  a non-relativistic superfluid are essentially different from 
the standard fluid equations. At non-vanishing temperature one has to employ the two-fluid description of Landau
~\cite{landaufluids}, which takes into account  the motion of both the superfluid and of the normal component of the system.   
In order to describe the different dissipative processes one has to introduce more transport coefficients
than in a normal fluid. In particular, one has three independent bulk viscosities, as well as the  shear viscosity and the thermal conductivity. 
The same occurs in the relativistic version of Landau's two fluid model, as we briefly review here (see Refs.~\cite{Son:2000ht,Gusakov:2007px}
for more explicit details).

In this paper we will follow the approach derived by Son \cite{Son:2000ht}. This formulation takes into account that
superfluidity arises after the appearance of a quantum condensate that spontaneously breaks a
$U(1)$ global symmetry. The Goldstone mode  associated with this breaking then  should appear in any low energy (and low momentum)
effective theory. Hydrodynamics, being a long spacetime dynamical formulation,  should then necessarily incorporate the Goldstone field in their equations.

The hydrodynamical  equations for the superfluid take the form of conservation laws
for both the current, $n^\mu$, and energy-momentum tensor, $T^{\mu \nu}$,  of the system
\begin{equation}
\partial_\mu n^\mu_q = 0 \ , \qquad \partial_\mu T^{\mu \nu} = 0 \ .
\end{equation}
One further adds the Josephson equation, which describes the dynamical evolution of
the Goldstone field, $\varphi$, or phase of the condensate
\begin{equation}
u^\mu \partial_\mu \varphi+ \mu = 0  \  ,
\end{equation}
 where $u^\mu$ is the hydrodynamical velocity, and $\mu$ is the chemical potential of the system. The 
stress-energy tensor  and the current are expressed
 as
 \begin{eqnarray}
 T^{\mu \nu} & = & \left(\epsilon_0 + P\right)u^\mu u^\nu -
P g^{\mu \nu}
+ V^2 \partial^\mu \varphi \,\partial^\nu \varphi  \  ,\\
 n^\mu_q & = & n_{q,0} u^\sigma - V^2 \partial^\mu  \varphi \  ,
\end{eqnarray}
where  $\epsilon_0$ stands for the energy density, $P$ is the pressure, and $V$ is a variable proportional
to the quantum condensate. The energy density obeys the relation
 \begin{equation}
 \label{pressure} 
 \epsilon_0= ST + n_{q,0} \, \mu - P \ ,
 \end{equation}
where $S$ is the entropy of the system.

Sometimes it is better to write  the hydrodynamical equations in terms of the new
 variable
\begin{equation}
w^\mu = - \left(\partial^\mu \varphi + \mu u^\mu \right) \ ,
\end{equation}
which represents a counterflow momentum, see Ref.~\cite{Gusakov:2007px}. In the non-relativistic limit, the spatial component of this four vector
corresponds to the counterflow momentum of Landau's two-fluid model formulation ~\cite{Gusakov:2007px}.

In the absence of dissipation it is possible to show that the entropy current is conserved,
$\partial_\mu (S u^\mu) = 0$. There is generation of entropy, otherwise. 
The dissipative terms associated to the above hydrodynamical equations have been constructed in Ref.~\cite{Gusakov:2007px},
showing that in the non-relativistic limit they correspond to those appearing in Landau's two-fluid model. 
After defining a comoving frame, where $u_\mu = (1,0,0,0)$  one can get the terms associated
with all the viscosity coefficients and thermal conductivity.

In this article we will only be interested in the three bulk viscosity coefficients, which modify the equilibrium hydrodynamical equations.
  For purposes of interpretation of these coefficients, and for the explicit computations in
the remaining part of this article, we can see them as follows.
 The friction forces due to the bulk viscosities can be understood as modifications, with respect to their equilibrium values, in the main driving forces acting on the normal and 
superfluid components. These forces are given by the gradients of the pressure $P$ and the chemical potential $\mu$, respectively. One can write 
\ba
\label{Inter-bulksP}
P &=& \bar P + \delta P =  \bar  P  - \zeta_1 \,\partial_\mu (V^2 w^\mu) -\zeta_2 \, \partial_\mu u^\mu\,, \\
\label{Inter-bulksmu}
\mu &=& \bar \mu + \delta \mu =  \bar \mu
- \zeta_3 \, \partial_\mu (V^2 w^\mu)  -\zeta_4 \,\partial_\mu u^\mu \, ,
\ea 
where $\bar P$ and $\bar \mu$ are the equilibrium pressure and chemical potential, respectively.  
According to the Onsager symmetry principle, the coefficients should satisfy $\zeta_1 = \zeta_4$, so that in
fact there are only three independent coefficients.
It is also important to stress that the requirement  of positive entropy production imposes that 
$\zeta_2, \zeta_3$   should be positive and that  $\zeta_ 1^2 \leq \zeta_2 \zeta_3$   ~\cite{Gusakov:2007px}.
 While $\zeta_2$ has the same meaning as the one that occurs in a normal fluid, $\zeta_1$ and $\zeta_3$
refer to dissipation that occurs  when the superfluid counterflow is not incompressible.

The expressions given in Eqs.~(\ref{Inter-bulksP},\ref{Inter-bulksmu}) are interesting, as they show us that  in order to
compute the different viscosity coefficients we will simply have to evaluate the modifications
of both the pressure and chemical potential when the system leaves the equilibrium configuration. 
This is what we will do in the remaining part of this article for the CFL phase, using this hydrodynamical
formulation, where $n_q^\mu$ is the quark current, and $\mu$ the quark chemical potential. In that case
the parameter $V^2$ can be computed using a Ginzburg-Landau formalism, see Ref.~\cite{Iida:2002ev} for the computation
of $V$ for the CFL superfluid.

\section{Particle number changing processes to restore equilibrium after a compression or rarefaction}
\label{decay-sec}

After an expansion (or rarefaction) of a system in equilibrium, the pressure diminishes (or increases above)
its equilibrium value. Microscopic processes that involve reactions that change the particle densities take place to
restore the equilibrium configuration. These kind of processes are needed as  the system 
cannot equilibrate  to a different value of the
volume without modifying the density of particles.

 We consider CFL matter at relatively low temperature, where $T \ll T_c$, where
$T_c$ is the critical temperature for the onset of superfluidity, which is believed to be above several tens of MeV.
 In this temperature regime we expect that
all the transport coefficients are dominated by the lightest degrees of freedom of CFL quark matter, as
the contribution of the heavier modes is Boltzmann suppressed. The lightest modes in the CFL phase
are the superfluid phonon, which is the Goldstone mode associated with the spontaneous breaking of $U(1)_B$
and remains always massless, and the kaons, which are pseudo Goldstone modes associated with
chiral symmetry breaking. Effective field theories associated to these degrees of freedom can be constructed from
QCD \cite{Casalbuoni:1999wu,Son:2000cm,Bedaque:2001je,Son:2002zn,Anglani:2011cw}.

In this Section we review the most relevant reaction rates of particle number  changing processes that involve the
CFL light degrees of freedom, and which are relevant for the computation of the bulk viscosity coefficients \cite{Mannarelli:2009ia,Alford:2007rw}.

{\bf 1.  Emission, and absorption, of phonons by phonons.}
 The superfluid phonons are  massless degrees of freedom. Their dispersion law is of the form
 \begin{equation}
 E_p = c_s p + B p^3 + \cdots \ ,
 \end{equation}
 where $c_s$ is the speed of sound, and $B$ is also a density dependent parameter that in the CFL phase obeys $B <0$ ~\cite{Zarembo:2000pj}. 
 As realized in Ref.~\cite{Mannarelli:2009ia}, and because $B < 0$, the first possible phonon number changing  reaction involves
 five phonons $ \varphi +\varphi \longleftrightarrow \varphi+\varphi +\varphi $, as the so-called Beliaev process 
 that describes the decay of one phonon in two is only kinematically possible for a dispersion law with positive
 values of $B$.

 It has been also found that the decay rate associated to the five phonon reaction  behaves very differently
 for large or small angle scatterings, being the last processes those which dominate. The decay associated to
 five phonon small scatterings has not been computed, but power counting techniques allows one to estimate
 it as behaving as 
 \begin{equation}
 \label{ph-decay}
\Gamma_{ph} = a_{ph} \,g_3^2\,g_4^2 \frac{T^{12}}{c_s^6B^2} \ ,
\end{equation}
  where $g_3$ and $g_4 $ are the phonon self-coupling constants
associated to the three-phonon and four-phonon vertices, respectively, and   $a_{ph}$ is  an unknown number.

Taking into account the value of the strange quark mass, $m_s$, one can find the value of all the phonon self-coupling parameters 
 to leading order in a $m_s^2/\mu^2$  expansion. In the
density of interest for compact stars, this is the typical small
parameter used to expand our knowledge of CFL at asymptotic
densities. In particular, one finds 
\cite{Manuel:2007pz}
\begin{equation}
\label{m-speedsound}
c_s^2  =  \frac 13 \left( 1 - \frac{m^2_s}{3 \mu^2} \right) \ ,  \qquad
g_3  =   \frac {\pi}{9 \mu^2} \left( 1 + \frac{m^2_s}{4 \mu^2} \right) \ , \qquad
g_4  =    \frac {\pi^4}{108 \mu^4} \left( 1 + \frac{m^2_s}{3 \mu^2} \right) \, .
\end{equation}

Furthermore, a microscopic computation with vanishing quark masses allows one to find the phonon dispersion law at cubic order  ~\cite{Zarembo:2000pj} 
\begin{equation}\label{newB}
B = -\frac{11 c_s}{540 \Delta^2}  \, .
\end{equation}

{\bf 2.  Decay of neutral kaons (and antikaons) into two superfluid phonons}. This is a process mediated
by electroweak interactions, which allow for this flavor-changing reaction. First, a kaon is converted into
a virtual phonon, which  finally  decays into two other superfluid phonons, so the total reaction
can be seen as $ K^0 (\bar K^0) \longleftrightarrow \varphi+\varphi$.
The neutral kaons/antikaons have a dispersion law of the form
\begin{equation}
E_{K^0/\bar K^0} = \mp \mu_{K^0}^{eff} + \sqrt{ v^2 p^2 + m_{K^0}^2} \ ,
\end{equation}
where $v = \frac 13$,  and $\mu_{K^0}^{eff}$ is an effective
chemical potential given by 
\begin{equation}
\mu_{K^0}^{eff} = \frac{m_s^2 - m_d^2}{2 \mu} \ ,
\end{equation}
where $m_d$ is the down quark mass, that in this work we will take as negligible in front of the strange quark mass.
The neutral kaon mass  $m_{K^0}$ was computed in the limit of asymptotic high density \cite{Son:2000cm,Bedaque:2001je}. However, at more
moderate values  instantons effects, which are not under full control, give a contribution to all the meson masses \cite{Manuel:2000wm,Schafer:2002ty}.
As a result, the kaon mass is poorly known, and we will take it here as an unknown parameter.

The decay rate associated with this process  has been computed in Ref.~\cite{Alford:2007rw}, where it has been shown
that in a good range of temperatures it can be approximated by
\begin{eqnarray}
\label{decayrate1}
\Gamma_{K,ph}
\approx
\frac{G_{ds}^2 f_{\pi}^2 f_H^2}{18 \sqrt{3} \pi}
(1+\frac{m^2_{K^0}}{\mu_{K^0}^{eff,2}} )\bar{p}^4
\frac{e^{v\bar{p}/T}}{(e^{v\bar{p}/T}-1)^2}\,,
\end{eqnarray}
where the low energy effective field theory coupling constants read
\begin{equation}
  f_H^2 = \frac{3}{4} \left( \frac{\mu^2}{2 \pi^2} \right) \ ,  \qquad
  f_{\pi}^2  = \frac{21 - 8 \log(2)}{18} \frac{\mu^2}{2 \pi^2}   \ ,  \qquad
  G_{ds}  =  \sqrt{2}\,V_{ud}V_{us} G_F \approx  0.304 \, G_F\
  \end{equation}
 where $G_F$ is the Fermi constant and   
\begin{equation}
\bar{p} = \frac{ m_{K^0}^2 - \mu_{K^0}^{eff, 2}}{2 v \mu_{K^0}^{eff}}  
\end{equation}
is the momentum where the  virtual phonon that mediates the  reaction is
nearly on-shell. This expression becomes invalid for very low temperatures,
$T \ll T_a = \frac{\delta m^2}{\mu_{K^0}^{eff}}$, or for large values of $T$, which are
outside the value of physical interest  ~\cite{Alford:2007rw}. In our numerical computations
for the bulk viscosities we will use this approximated form of the kaon decay rate, as in the low
$T$ regime where this expression is not valid  the kaon decay rate becomes meaningless for
the computation of the frequency dependent bulk viscosities (see Table 2 of Ref. ~\cite{Alford:2007rw}).

Other processes that involve the kaons are very much suppressed with respect to the one described above.
In particular, it is possible to show that the rate of the reaction $ K^0 + \varphi \longleftrightarrow \varphi$
is suppressed with respect to the decay rate considered above. 
In Ref.~\cite{Alford:2007rw}, it has been also shown that this is the case 
for the reaction rates involving charged kaons and leptons.

\section{Phonon and Kaon contribution to the bulk viscosity coefficients}
\label{bulk-sec}

In this Section we will compute the phonon and kaon contribution to the bulk viscosities following the same strategy
devised by Khalatnikov to compute the bulk viscosities coefficients due to phonons and rotons in He$^4$. 
The computation is quite analogous, even in  the CFL superfluid  the kaons are collective modes with different quantum numbers than the phonons, 
while in He$^4$ phonons and rotons are collective modes with the same quantum numbers : they are actually the same
excitations but seen in a different momentum regime. In any case, all what matters in the computation is to have two well-defined different collective modes which
interact with each other.

The superfluid phonon and  the kaon are bosons. At equilibrium the number
densities of these particles are entirely determined by the temperature 
$T$ and by  their dispersion laws, which depend on the quark chemical potential, and on
QCD parameters, such as the gauge coupling constant and quark masses.
 Small deviations from equilibrium can be parametrized with some ``fake" phonon and
kaon  chemical potentials,  $\delta \mu_{\rm ph}$  and $\delta \mu_k$, respectively \footnote{As in Ref.~\cite{Alford:2007rw} we will consider that the effective kaon chemical
potential $\mu_{K^0}^{eff}$ is part of its dispersion law, that is, a parameter that is fixed by the value of
the quark chemical potential}.  Equilibrium is restored by microscopic processes that modify the particle
number densities. The dynamical evolution of the phonon and kaon particle densities will be
governed by the following equations
\begin{eqnarray}
\label{mu1}
 \partial_\mu(n_{ph}\,u^{\mu})& =&- \gamma_{ph ph} \,  \delta\mu_{ph}
 -\gamma_{ph k}\, \delta\mu_{k}\,,
\\
\label{mu2}
 \partial_\mu(n_k\,u^{\mu})& =&-\gamma_{kk}\,\delta\mu_{k}
 -\gamma_{k ph}\,\delta\mu_{ph}\,,
\end{eqnarray}
where the different $\gamma_{ij}$ parameters are kinetic coefficients which have to be symmetric in their indices, and
which can be related to the decay rates  associated to the microscopic processes under consideration \cite{IntroSupe}.
If we take into account only the scattering processes mentioned in the previous Section one then has
\begin{equation}
\gamma_{ph ph} = \frac{\Gamma_{ph} + 4 \, \Gamma_{K, ph}}{T} \ , \qquad \gamma_{ph k} =\gamma_{ k ph} =
 -2 \, \frac{ \Gamma_{K,ph}}{T} \ ,  \qquad  \gamma_{kk} = \frac{ \Gamma_{K,ph}}{T}  \ .
 \end{equation}

As emphasized before, the equilibrium phonon and kaon densities are functions of $\mu$ and $T$, or equivalently, $n_q$ and $S$. 
For small departures from equilibrium, one can follow the change in the phonon and kaon number densities by following the
dynamical evolution of $n_q$ and $S$. To this end, we split  all thermodynamical and hydrodynamical variables   into their equilibrium values, denoted with a bar,  and a small fluctuation,
denoted with a $\delta$. Thus, we write
\begin{equation}
 n_{ph} = \bar n_{ph} + \delta n_{ph} \ ,
\end{equation}
and similarly for the remaining thermodynamic and hydrodynamical variables.
We then write
\begin{equation}
 \delta n_{ph}=\frac{\partial n_{ph}}{\partial n_q}\delta{n}_q
	+ \frac{\partial n_{ph}}{\partial S}\delta S\ , \qquad  
	\delta n_{k}=\frac{\partial n_{k}}{\partial n_q}\delta{n}_q
	+ \frac{\partial n_{k}}{\partial S}\delta S \ ,
	\end{equation}
where the derivatives are taken at the equilibrium values.	
Because the  deviation out of equilibrium is assumed to be sufficiently slow
we can use the linearized hydrodynamical equations to evaluate the dynamical evolution
of $\delta n_q$ and $\delta S$ \footnote{Note that in order of not overcharging the notation, we avoid the use
of expressions like $\partial_\mu \delta u^\mu$.}
\begin{eqnarray}
\bar u^\mu \partial_\mu \delta n_q & = & - \bar{ n}_q \, \partial_\mu u^\mu - \partial_\mu (V^2 w^\mu) \ , \\
\bar u^\mu \partial_\mu \delta S & = & - \bar {S} \, \partial_\mu u^\mu \ ,
\end{eqnarray}
to extract the value of the phonon and kaon chemical potentials. One finds
\begin{eqnarray}
\delta \mu_{ph} & =& \frac{\gamma_{kk}}{\gamma_{k ph}^2  - \gamma_{kk} \gamma_{ph ph}} \left( I_2^{ph} \partial_\mu u^\mu - I_1^{ph} \partial_\mu (V^2 w^\mu) 
\right) - \frac{\gamma_{k ph}}{\gamma_{k ph}^2  - \gamma_{kk} \gamma_{ph ph}} \left( I_2^{K} \partial_\mu u^\mu - I_1^{K} \partial_\mu (V^2 w^\mu)  
\right) \ ,
\\
\delta \mu_{k} & =& -\frac{\gamma_{k ph}}{\gamma_{k ph}^2  - \gamma_{kk} \gamma_{ph ph}}
 \left( I_2^{ph} \partial_\mu u^\mu - I_1^{ph} \partial_\mu (V^2 w^\mu)
\right) + \frac{\gamma_{ph ph}}{\gamma_{k ph}^2  - \gamma_{kk} \gamma_{ph ph}}
 \left( I_2^{k} \partial_\mu u^\mu - I_1^{k} \partial_\mu (V^2 w^\mu)  
\right) \ ,
\end{eqnarray}
where we have defined  the functions 
\begin{equation}
\label{I-functions}
I_1^i  \equiv   \frac{\partial n_{i}}{\partial {n_q}}\ ,  \qquad
 I_2^i  \equiv   \bar n_{i}- \bar S \,\frac{\partial n_{i}}{\partial S}
	-\bar{n}_q \, \frac{\partial n_{i}}{\partial n_q}
 \ , \qquad i = {\rm ph}, k  \ .
\end{equation}

Finally, we compute the change in both the pressure and quark chemical potential caused by the presence of a non-vanishing $\delta \mu_k$ and $\delta \mu_{ph}$
\begin{eqnarray}
\label{varP}
\delta P & = & \frac{\partial P}{\partial \mu_{\rm ph}} \delta \mu_{\rm ph} + \frac{\partial P}{\partial \mu_{k}}\delta \mu_{k}   \ , \\ 
\label{varmu}
\delta \mu & = & \frac{\partial \mu}{\partial \mu_{\rm ph}} \delta \mu_{\rm ph} + \frac{\partial \mu}{\partial \mu_{k} }\delta \mu_{k}  \ .
\end{eqnarray}
Using the thermodynamical relation Eq.~(\ref{pressure}) and  the identity \cite{IntroSupe}
\begin{equation}
d \epsilon_0 = T dS + \mu dn_0 -n_{ph} d \mu_{ph} - n_k d \mu_k \ ,
\end{equation}
one finds
\begin{equation}
I_1^i  = - \frac{\partial \mu}{\partial \mu_i}   \ , \qquad
 I_2^i  =  \frac{\partial P}{\partial \mu_i} 
 \ , \qquad i = {\rm ph}, k  \ .
\end{equation}

The bulk viscosity coefficients can then be obtained by identifying the different coefficients in the variations of
the pressure
 and chemical potential that are proportional to the divergences of the two velocity vectors, using
 Eqs~(\ref{Inter-bulksP},\ref{Inter-bulksmu}). One then finds
\begin{eqnarray}
\zeta_1 &=&
\label{staticbulks1}
-\frac{T}{\Gamma_{ph}} \left(I_1^{ph} + 2 I_1^k \right) \left(I_2^{ph} + 2 I_2^k \right)
-\frac{T}{\Gamma_{K,ph}}I^k_1I^k_2
\,,
\\
\label{staticbulks2}
\zeta_2&=&
\frac{T}{\Gamma_{ph}}  \left(I_2^{ph} + 2 I_2^k \right) \left(I_2^{ph} + 2 I_2^k \right)
+\frac{T}{\Gamma_{K,ph}}(I^k_2)^2\,,
\\
\label{staticbulks3}
 \zeta_3&=&
\frac{T}{\Gamma_{ph}}  \left(I_1^{ph} + 2 I_1^k \right) \left(I_1^{ph} + 2 I_1^k \right)
+\frac{T}{\Gamma_{K,ph}}(I^k_1)^2
\ , \\
\zeta_4 & =& \zeta_1 \ .
\end{eqnarray}
We note that the last equality is related to the symmetry properties of the kinetic coefficient $\gamma_{k ph}$.
If the symmetry on the indices of $\gamma_{ij}$ is violated, then the Onsager symmetry
principle is not fullfilled, and then $\zeta_1 \neq \zeta_4$.

As emphasized at the beginning of this Section, we  want to stress that the expressions we found  for the viscosities
 are very similar to those found for the phonon and roton contributions to the bulk viscosities to $^4$He
 (see Ref.~\cite{IntroSupe}).

In the situation where the temperature is much smaller than the kaon gap $T \ll \delta m = m_k^0 - \mu_{K^0}^{\rm eff}$, then the contribution of the
kaons is thermally suppressed. In  Ref.~\cite{Mannarelli:2009ia} the three bulk viscosity coefficients due to the phonons were evaluated, that is,
in the situation where $I^k_i =0$.
Because the kaon gaps are poorly known, it is not clear at which values of $T$ the kaons might become relevant or not for transport in the CFL phase.

While the phonons are never thermally suppressed because they are massless degrees of freedom, one can still find a regime where the kaons dominate
the contribution to the bulk viscosities. This may happen when $\Gamma_{ph} \gg \Gamma_{K,ph}$, as then the first terms of
Eqs.~(\ref{staticbulks1}-\ref{staticbulks3}) would be suppressed. Because of the high power dependence of the
phonon decay rate, one may expect that this is the situation that one may encounter at high temperatures.

One   important lesson deduced from the expressions written above is that in the domain where the kaons are not thermally suppressed the pure
phonon and kaon contributions cannot be disentangled, as the bulk viscosities  contain products of  both phonon and kaon 
thermodynamical functions. That is, the total bulk viscosity in the presence of phonons and kaons is not simply the sum of the phonon and kaon
contributions.

\section{Phonon and kaon contributions to the frequency-dependent bulk viscosities}


For astrophysical applications it is important to find the value of the bulk viscosities when there is perturbation which is periodic in time, with an angular
frequency $\omega_c$. Then, one also assumes that the time evolution of all the physical variables is also periodic. The bulk viscosities then
turn out to be also a function  of $\omega_c$, and in this Section we will compute the frequency-dependent value of these transport coefficients.
In Subsec.~\ref{algebraic} we present the algebraic expressions of the viscosities, and we leave the presentation of the numerical results
for Subsec.~\ref{numerical}.

\subsection{Analytical results}
\label{algebraic}

The evaluation of the frequency dependent bulk viscosities when only phonons are considered was done in Ref.~\cite{Mannarelli:2009ia}.
In Ref.~\cite{Alford:2007rw}, $\zeta_2 (\omega_c)$ due to kaons was also computed. Here we want to get the values of the frequency dependent viscosities when
both the phonons and the kaons are taken into account. It is not a priori obvious how the coefficients will behave, as we have already seen that in
a non-periodic situation one cannot simply sum the phonon and kaon contributions to the bulk viscosities.

We will work in the comoving frame, where $\bar u^\mu = (1,0,0,0)$. Furthermore, as in Ref.~\cite{Alford:2007rw} we will compute the viscosities assuming
that the temperature is constant. In this regard, our computation here differs from that carried out for the phonons in Ref.~\cite{Mannarelli:2009ia}, where
such an assumption was not considered.

For purposes of comparison with the results of a non-periodic case presented in the previous Section, and the results obtained in Ref.~\cite{Alford:2007rw},
we will first  find the viscosities
 neglecting the variation of the phonon number, that is, we will assume momentarily that $\delta n_{ph} = \delta \mu_{ph} =0$.
We thus use only the kaon and quark density equations. In the comoving frame, and taking into account the explicit time dependence of the variables,
one has
\begin{eqnarray}
&i \omega_c \, \delta n_{k} & + \bar n_{k} \, {\rm div} {\bf u} = - \gamma_{kk}\, \delta \mu_{k}  \ , \\
& i \omega_c \, \delta n_{q} & + \bar n_{q} \, {\rm div} {\bf u} +{\rm div} (V^2 {\bf w}) = 0 \ .
\end{eqnarray}
We  now express
 the fluctuations of the particle densities in terms  of a change in both the quark and kaon chemical potentials
\begin{eqnarray}
\delta n_q & = & \frac{\partial n_q }{\partial \mu} \, \delta \mu +\frac{\partial n_q }{\partial \mu_k} \,\delta \mu_k   \ ,
\\
\delta n_k & =& \frac{\partial n_k }{\partial \mu} \, \delta \mu +\frac{\partial n_k }{\partial \mu_k} \, \delta \mu_k  \  ,
\end{eqnarray}
where the derivatives of the particle number densities are evaluated in equilibrium, where one has $\delta \mu_k =0$,
see Appendix ~\ref{freenergy} for explicit details.
In this way, we can solve for $\delta \mu$ and $\delta \mu_k$ as a function of  the different thermodynamical variables
and the angular frequency. The solutions, and therefore also the bulk viscosities, are complex quantities.
However, for the computation of the dissipation in energy only the real part of these coefficients is needed, so we will only
 extract the real part of the viscosities.

Using Eqs.~(\ref{Inter-bulksP},\ref{Inter-bulksmu}) and  Eqs.~ (\ref{varP},\ref{varmu}), we find that the real part of the viscosities can be expressed as
 \begin{equation}
 \label{kaonwbulk}
\zeta_i(\omega_c)=  \frac{\gamma_{eff}}{\omega_c^2 + \gamma_{eff}^2} C_i \ ,
\end{equation}
where
\begin{eqnarray}
C_1 &=& C_4 = - \frac{\frac{\partial n_k}{\partial \mu}}{ \frac{\partial n_q }{\partial \mu} } \frac{1}{L} \left( \bar n_k \frac{\partial n_q }{\partial \mu} - \bar n_q \frac{\partial n_k }{\partial \mu} \right) \ ,
 \\
C_2 & = & \frac{1}{ \frac{\partial n_q }{\partial \mu} } \frac{1}{L} \left( \bar n_k \frac{\partial n_q }{\partial \mu} - \bar n_q \frac{\partial n_k }{\partial \mu} \right)^2 \ , \\
C_3 & = & \frac{1}{ \frac{\partial n_q }{\partial \mu} } \frac{1}{L}  \left(\frac{\partial n_k}{\partial \mu}\right)^2  \ ,
\end{eqnarray}
and
\begin{equation}
 \gamma_{eff} = \frac{\Gamma_{Kph}}{T}  \frac{\partial n_q }{\partial \mu}  \frac{1}{L} \ ,
\end{equation}
and we have defined the variable
\begin{equation}
L \equiv  \frac{\partial n_q }{\partial \mu} \frac{\partial n_k }{\partial \mu_k} - \frac{\partial n_q }{\partial \mu_k} \frac{\partial n_k }{\partial \mu} \ .
\end{equation}

It is interesting to note that the condition for positive entropy production is saturated, and we have $\zeta_1^2 = \zeta_2 \zeta_3$.

With the results presented above, we can check that the formal expression of  $\zeta_2$ agrees with the bulk viscosity computed  in  Ref.~\cite{Alford:2007rw},
as one has $\frac{\partial n_k}{\partial \mu} = \frac{\partial n_q}{\partial \mu_k}$ (see Appendix ~\ref{freenergy}). We can also check that
in the limit  $\omega_c  \rightarrow 0$  we obtain the values of the bulk viscosities obtained in the previous Section, in the situation where one assumes no
thermal or entropy variation, and for $I_1^{ph} = I_2^{ph} = 0$.
In order to check this limit, one simply has to realize that at constant temperature
\begin{equation}
 \frac{\partial n_k}{\partial \mu} \left(   \frac{\partial n_q}{\partial \mu } \right)^{-1} =  \frac{\partial n_k}{\partial n_q} = I^k_1  \ , 
 \end{equation}
to write down all the pieces $C_i/ \gamma_{eff}$ in terms of the decay rate and also the
 $I_1^k$ and $I_2^k$ functions introduced in Eq.~(\ref{I-functions}).

Let us consider now the presence of fluctuations in the phonon sector. 
The set of linear equations that we have to solve should now also involve the phonon number density equation. The system of equations to solve is
\begin{eqnarray}
&i \omega_c \, \delta n_{ph} &+ \bar n_{ph} \, {\rm div} {\bf u} = - \gamma_{ph ph} \, \delta \mu_{ph} - \gamma_{ph k} \, \delta \mu_k  \ , \\
&i \omega_c \, \delta n_{k} & + \bar n_{k} \,{\rm div} {\bf u} = - \gamma_{kk} \, \delta \mu_{k} - \gamma_{k ph} \, \delta \mu_{ph}  \ , \\
& i \omega_c \,  \delta n_{q} & + \bar n_{q} \, {\rm div} {\bf u} +{\rm div} (V^2 {\bf w}) = 0 \ .
\end{eqnarray}
One can find solutions for $\delta \mu$, $\delta \mu_k$ and $\delta \mu_{ph}$ if  we  express the fluctuations of the particle densities in terms of the fluctuations of the chemical potentials
\begin{eqnarray}
\delta n_q & = & \frac{\partial n_q }{\partial \mu} \delta \mu + \frac{\partial n_q }{\partial \mu_k} \delta \mu_k +\frac{\partial n_q }{\partial \mu_{ph}} \delta \mu_{ph}  \ ,
\\
\delta n_k & = &\frac{\partial n_k }{\partial \mu} \delta \mu +\frac{\partial n_k }{\partial \mu_k} \delta \mu_k  \ , 
\\
\delta n_{ph} & = & \frac{\partial n_{ph}}{\partial \mu} \delta \mu 
 +\frac{\partial n_{ph} }{\partial \mu_{ph}} \delta \mu_{ph}  \ .
\end{eqnarray}
Then,  using Eqs.~(\ref{Inter-bulksP},\ref{Inter-bulksmu}) and  Eqs.~(\ref{varP},\ref{varmu}), one can find  the real part of the viscosities. 
In this case we find that  the frequency dependence of the coefficients  is different from the case where only the kaons (or phonons) are considered. 
The real part of the three coefficients can now be written as 
 \begin{equation}
 \label{fullbulks}
\zeta_i(\omega_c)= \frac{\tilde D_i(\omega_c)}  {\omega_c^2 + \left (\frac{O+ R \omega_c^2}{Q} \right)^2}  \ ,
\end{equation}
where
\begin{eqnarray}
O & = & \frac{\Gamma_{Kph} \Gamma_{ph}}{T^2}  \left(\frac{\partial n_q }{\partial \mu} \right)^2  \ , \\
R & = & \frac{\partial n_q }{\partial \mu}  \left( - \frac{\partial n_{ph}}{\partial \mu_{ph}} L + \left( \frac{\partial n_{ph}}{\partial \mu} \right)^2 \frac{\partial n_k}{\partial \mu_k} 
 \right) \ ,\\
 Q  & =&  \frac{\partial n_q }{\partial \mu} \left\{ \left(\frac{\partial n_q }{\partial \mu}  \left( 4  \frac{\partial n_k }{\partial \mu_k} +  \frac{\partial n_{ph} }{\partial \mu_{ph} }
  \right) - \left(2  \frac{\partial n_k }{\partial \mu}  +  \frac{\partial n_{ph} }{\partial \mu} \right) \left(2  \frac{\partial n_q }{\partial \mu_k}  +  \frac{\partial n_{ph} }{\partial \mu} \right) \right) 
 \frac{ \Gamma_{K ph}}{T} + L \frac{ \Gamma_{ph}}{T} \right\} \ ,
 \end{eqnarray}
and $\tilde D_i$ are polynomials of second order in $\omega_c$, expressed as
\begin{eqnarray}
\tilde D_i (\omega_c)=\frac{1}{Q^2} \left(
b_i\,O+\omega_c^2\,(b_i\,R+a_i\,Q)
\right) \, .
\end{eqnarray}
The coefficients $a_i,$ and $b_i$  are different for each bulk viscosity, and depend on both the phonon and kaon thermodynamical variables, and on the the decay rates.
More explicitly, the coefficients for the first bulk viscosity are
\begin{eqnarray}
a_1&=&
-       \frac{\partial n_{ph}}{\partial\mu_{ph}} \frac{\partial n_k}{\partial\mu}   \left( \bar  n_k \frac{\partial n_q}{\partial\mu}   -  \bar n_q \frac{\partial n_k}{\partial\mu}  \right)
-  I_2^{ph}\,\frac{\partial n_{ph}}{\partial\mu}   \,  \frac{\partial n_k}{\partial\mu_k}  \,  \frac{\partial n_q}{\partial\mu}   
\,\,,\\
b_1&=& -\left(   \bar  n_k\, \frac{\partial n_q}{\partial\mu}- \bar n_q\, \frac{\partial n_k}{\partial\mu}   \right)
\left\{
  \frac{\Gamma_{ph}}{T}  \, \frac{\partial n_k}{\partial\mu}  
+
\frac{\Gamma_{K,ph}}{T}\,
\left(2\,  \frac{\partial n_{ph}}{\partial\mu}   +4\,  \frac{\partial n_k}{\partial\mu}   
\right)
\right\}
\\ \nonumber
&-&  I_2^{ph}\,
\frac{\Gamma_{K,ph}}{T}\,\frac{\partial n_q}{\partial\mu}   \,
\left(  \frac{\partial n_{ph}}{\partial\mu}   +2\,  \frac{\partial n_k}{\partial\mu}   
\right)
\ .
\end{eqnarray}
The coefficients for the second and third bulk viscosities are
\begin{eqnarray}
a_2&=&
\left( \bar n_k\, \frac{\partial n_q}{\partial\mu}   - \bar  n_q\, \frac{\partial n_k}{\partial\mu}   \right)
\left\{
\bar  n_{ph}\,\frac{\partial n_k}{\partial\mu}     \frac{\partial n_{ph}}{\partial\mu}
+ \bar n_k \left(  \frac{\partial n_q}{\partial\mu}     \frac{\partial n_{ph}}{\partial\mu_{ph}}   -  \frac{\partial n_{ph}}{\partial\mu} \,\frac{\partial n_q}{\partial\mu_{ph}}
\right)
- \bar n_q\, \frac{\partial n_k}{\partial\mu}     \frac{\partial n_{ph}}{\partial\mu_{ph}}  
\right\}\\
&+ & I_2^{ph} \frac{\partial n_q}{\partial\mu}   \left\{
\bar n_{ph}\,L
+  \bar n_k\,\frac{\partial n_{ph}}{\partial\mu}     \frac{\partial n_q}{\partial\mu_k}   
- \bar n_q\, \frac{\partial n_{ph}}{\partial\mu}     \frac{\partial n_k}{\partial\mu_k}   
\right\}\,,
\nonumber\\ 
b_2&=&
\frac{\Gamma_{K,ph}}{T}\left\{
\left( 2\, \bar n_k+ \bar n_{ph} \right)\,  \frac{\partial n_q}{\partial\mu}   
- \bar n_q
\left(2\,  \frac{\partial n_k}{\partial\mu}   +  \frac{\partial n_{ph}}{\partial\mu}   
\right)
\right\}
\\&\times&
\left\{
2\left(   \bar n_k\, \frac{\partial n_q}{\partial\mu} -  \bar  n_q\, \frac{\partial n_k}{\partial\mu} 
\right)+ I_2^{ph} \frac{\partial n_q}{\partial\mu}   
\right\}
+\frac{\Gamma_{ph}}{T}\left(
  \bar  n_k\,  \frac{\partial n_q}{\partial\mu}  -  \bar n_q\, \frac{\partial n_k}{\partial\mu} 
\right)^2\,,
\nonumber\\
\end{eqnarray}
and
\begin{eqnarray}
a_3&=&\left(
\frac{\partial n_{ph}}{\partial \mu}
\right)^2\,
\frac{\partial n_k}{\partial \mu_k}+
\left(\frac{\partial n_k}{\partial \mu} 
\right)^2\,
 \frac{\partial n_{ph}}{\partial \mu_{ph}}\,,\\
b_3&=& 
\frac{\Gamma_{Kph}}{T}\left(
4\,\frac{\partial n_k }{\partial \mu} \,
\left(\frac{\partial n_{ph} }{\partial \mu}
+\frac{\partial n_k }{\partial \mu} 
\right)
+\frac{\partial n_{ph} }{\partial \mu}\,\frac{\partial n_q}{\partial\mu_{ph}}
\right)
+\frac{\Gamma_{ph}}{T}\,
\left(
\frac{\partial n_k }{\partial \mu} 
\right)^2
\label{last3bulk}
\,\,,
\end{eqnarray}
respectively. Further, we have found $\zeta_1 = \zeta_4$.

 We have  checked that our expressions
in the  limit $\omega_c \rightarrow 0$ reproduce the results of the bulk viscosities in the limit when there are
no thermal or entropy fluctuations, obtained in the previous Section. 
While it is not easy to verify whether the expressions for the three bulk viscosities are consistent with the constraint of positive entropy production,
we have checked numerically that this is the case, as one always gets $\zeta_1^2 \leq \zeta_2 \zeta_3$.

Because of the rather non-trivial dependence of the expressions of  every bulk viscosity coefficient on the phonon and kaon thermodynamical variables
it is not easy to infer how the viscosities scale with the temperature,
or with other relevant scales of the problem, such as $\mu$.  It is also not easy to see when the phonons or the kaons give a subleading contribution.
A numerical analysis is then mandatory.  We present the numerical results of the viscosities in the following Subsection.

\subsection{Numerical results}
\label{numerical}

\begin{figure}[th!]
 \centering
  \includegraphics[clip=true,width=0.45\textwidth] {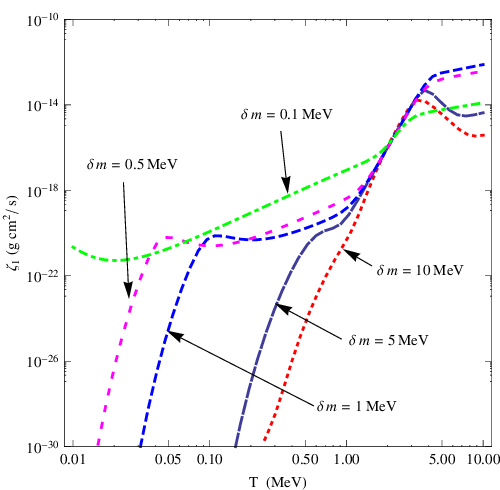} \hfill
  \includegraphics[clip=true,width=0.45\textwidth] {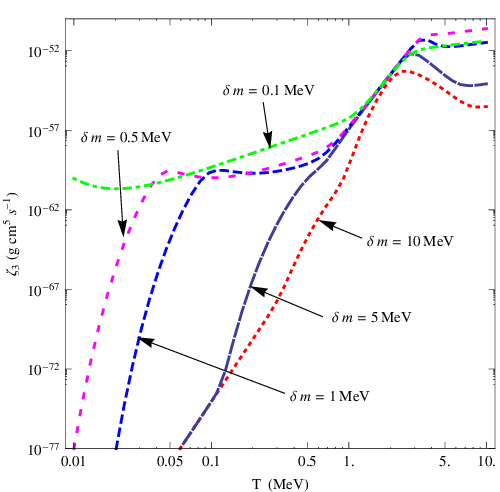}\\%
 \caption{Temperature dependence of the first and third bulk viscosities for the CFL quark matter phase due to thermal phonons and kaons for a frequency $\omega_c/2\pi = 1$ Khz. The phonon decay factor  is fixed at $a_{ph}=1$, and we
plot here the value of the viscosities for five different values of the kaon mass gap.}
 \label{fig:zeta1_total}
\end{figure}

\begin{figure}[th!]
 \centering
 \includegraphics[clip=true,width=0.45\textwidth] {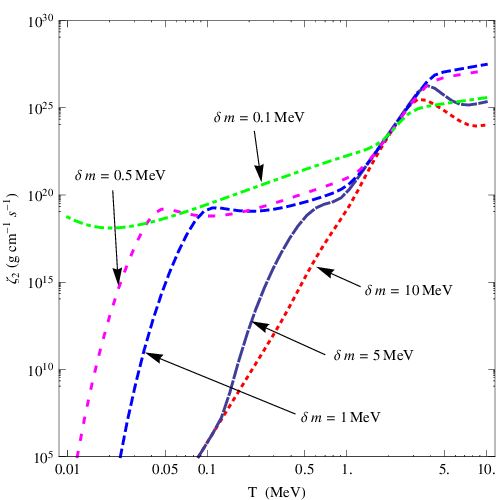} \hfill%
\includegraphics[clip=true,width=0.45\textwidth] {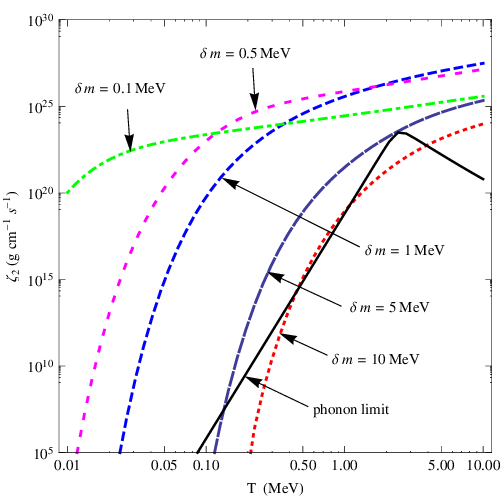}%
 \caption{Left: Temperature dependence of  second bulk viscosity for the CFL quark matter phase due to thermal phonons and kaons for a frequency $\omega_c/2\pi = 1$ Khz, with the phonon decay factor  is fixed at $a_{ph}=1$, and for five five different values of the kaon mass gap. Right:  The same plot but for $\zeta_2$ due to kaons only, or to phonons only. }
 \label{fig:zeta2_total}
\end{figure}

We have numerically computed the values of  the three bulk viscosities, and we present the results in different logarithmic plots for fixed values of the frequency, and as a function of the temperature. We consider temperatures in the range of $ 0.01\, {\rm MeV} < T < 10\, {\rm MeV}$. Lower temperatures might be relevant in astrophysical scenarios. However,
it was found out in Ref.~\cite{Manuel:2004iv} that for lower $T$ the phonon mean free path is much larger than the radius of a compact star, and a hydrodynamical description of the phonon fluid ceases to
be valid.

In our numerical computations we have  chosen the value of the quark chemical as $\mu = 400$ MeV,  the quark gap $\Delta = 25$ MeV 
and the strange quark mass $m_s=120$ MeV. We leave as free parameters the kaon gap $\delta m = m_{K^0} - \mu^{eff}_{K^0}$, and $a_{ph}$, the numerical factor present in the
phonon decay rate, see  Eq.~(\ref{ph-decay}). Notice that while we worked using natural units in the main part of the manuscript, we use SI units to show the numerical values of the viscosities. 
It is important to realize that the three bulk viscosities have different scalings in length (see the Appendix~\ref{dimensions}).

The phonon and kaon thermodynamical variables that are needed for the viscosities can be computed from the free energy of the system (see Appendix~\ref{freenergy}
for details), and we use the decay rates shown in Section~\ref{decay-sec}.

In Fig.~\ref{fig:zeta1_total} and Fig.~\ref{fig:zeta2_total} left
 we present the numerical values of $\zeta_1, \zeta_3$ and $\zeta_2$, respectively, for a fixed value of the frequency, $\omega_c/2 \pi= 1 $Khz, and for five different values of the kaon
 mass gap, ranging from $\delta m= 0.1$ MeV to $10$ MeV. We have fixed in these plots the value of  $a_{ph}=1$.
 Roughly speaking, the plots for the three viscosities present a similar shape, that is, a similar dependence on $T$, 
 although the numerical values for each viscosity are very different due to the different scaling dimensions of
each coefficient.

In Fig.~\ref{fig:zeta2_total} right we also present the value of   $\zeta_2$ when only the kaon contribution is taken into account, that is, with the results of Ref.~\cite{Alford:2007rw},
or when only the phonon contribution is taken into account (the results here differ from those of Ref.~\cite{Mannarelli:2009ia}  because here we are doing the computation assuming that the temperature remains constant).
The same plots for the kaon contribution only for  $\zeta_1$ and $\zeta_3$ present a rather similar shape, so we won't show them here.

\begin{figure}[th!]
 \includegraphics[clip=true,width=0.45\textwidth] {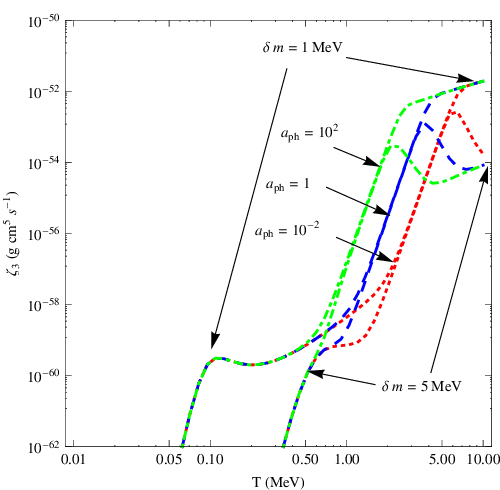}%
 \caption{Temperature dependence of the third bulk viscosity, for a fixed frequency $\omega_c = 1$ Khz,  for two different values of the kaon mass gap, and varying the value of
 the phonon decay factor from $a_{ph} = 10^{-2} - 10^2$.}
 \label{fig:zeta-aph_total}
\end{figure}

In Fig.~\ref{fig:zeta-aph_total}  we show the sensitivity of one of the bulk viscosities to the value of $a_{ph}$, the constant in the phonon decay rate Eq.~(\ref{ph-decay})
and for  two fixed values of the kaon mass gap. We can observe that the numerical values 
of the viscosities are rather insensitive to the values of $a_{ph}$ ranging form  $10^{-2}$ to $10^2$, except in the high $T$ limit. This is due to the high power
dependence on $T$ of the phonon decay rate, Eq.~(\ref{ph-decay}). For this reason, we have fixed the value of $a_{ph}=1$ in most of the plots we are showing.

\begin{figure}[th!]
 \centering
  \includegraphics[clip=true,width=0.45\textwidth] {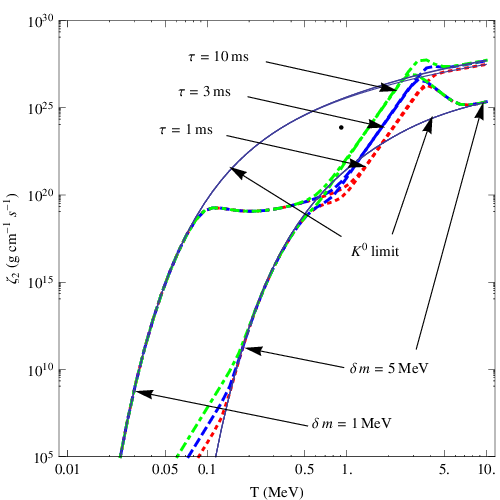}\\%
 \caption
{Temperature dependence of the second bulk viscosity for three different values of the frequency, taking $a_{ph} =1$, and for two different values of the kaon mass gap, For comparison
we have also plotte in the same graphs the values of $\zeta_2(\omega_c)$ when only the kaon contribution is taken into account.}
 \label{fig:zetaw_total}
\end{figure}

In Fig.~\ref{fig:zetaw_total} we show the value of $\zeta_2$ varying the value of the frequency, for $\omega_c = 1, 1/3, 0.1$ Khz, and we have shown also in the same
plot the same results for the kaon contribution only.

Some generic comments can be formulated from our numerical studies. At very small temperatures the phonons dominate the contribution to the bulk viscosities, as the
kaons are then thermally suppressed. This fact is not always shown clearly in the plots as this tends to occur at $T \ll 0.01$ MeV, which is outside the temperature regime where
one may expect the phonons to be in a hydrodynamical regime inside the star. At very large temperature $T \geq 10$ MeV the kaon contribution seems to dominate the behavior of the viscosities.
The reason comes from noting that in that regime $\Gamma_{ph}$ becomes very large, and suppresses the phonon contribution (see comments at the end of
Sec.~\ref{bulk-sec}). In the intermediate temperature regime we
always find an interval where first the kaons dominate, followed by a generic two-peak/hump structure, where both phonon and kaon contributions are relevant.
 The temperatures where this happens depends on the value of the kaon mass gap. 
 
One of the main conclusions derived from our numerical studies is that in the regime where the hydrodynamical description might be valid for the superfluid phonons, which should be
in the range $T  \geq 0.01$ MeV, and for a range of values of kaon gaps that do not exceed the few MeV, both kaons and phonons contribute to the value of the
viscosities. The superfluid phonon would only be highly dominant for  $T  \ll 0.01$ MeV, but as already mentioned, in this regime in a compact star the phonons do not collide
enough times to maintain a fluid description in that case. Then the phonons rather behave as a gas, and one should consider other dissipative effects in the system, such as that coming from the collisions
of phonons with superfluid vortices  \cite{Mannarelli:2008je}. 

In Ref.~\cite{Alford:2007rw} it was realized that the value of $\zeta_2(\omega_c)$ as due to kaons only is much less than the value of the same coefficient for unpaired quark matter. 
From our numerical results, we can say that if we include both the phonon and kaon contribution (see Fig.~\ref{fig:zeta2_total} and Fig.\ref{fig:zetaw_total}) the reduction is
still more severe, although this depends on the actual value of $\delta m$ and on the values of $T$.

\section{Outlook}
\label{conclu}


We have presented a computation of the frequency dependent bulk viscosities in the low temperature regime of CFL quark matter, extending previous results
already obtained in the literature.  The main contributions to the viscosities come from both kaons and superfluid phonons. The main results of this manuscript
 are presented in Eqs.~(\ref{fullbulks}-\ref{last3bulk}) and Figs.~\ref{fig:zeta1_total} to \ref{fig:zetaw_total}.

One of the main conclusions of our study is that in the regime where the hydrodynamical description might be valid for the superfluid phonons, which should be
in the range $T  \geq 0.01$ MeV, and for a range of values of kaon gaps do not exceed the few MeV, both kaons and phonons contribute to the value of the
viscosities, and one cannot discard any of the two quasiparticle contributions. Our results also agree with the main finding of  Ref.~\cite{Alford:2007rw}, that is,
that the coefficient   $\zeta_2$ in the CFL phase is several order of magnitude smaller than the same coefficient in unpaired quark matter.

The studies of transport coefficients are important in order to find possible signatures of quark matter in astrophysical scenarios. The fate of different set of
oscillations modes of  compact stars are governed by the value of the transport coefficients in the star. In particular, the fate of the  r-modes of the star \cite{Andersson:2000mf}
depend on the dissipative effects in the star. Previous analysis of the damping of the r-modes have been carried out in the literature 
\cite{Madsen:1999ci,Jaikumar:2008kh,Mannarelli:2008je,Andersson:2010sh}. The results of our study should be taken into account in a more refined study of
the the damping of r-modes, where the value of the three bulk viscosity coefficients should be relevant

\begin{acknowledgments}
We thank M. Mannarelli for useful discussions.
This work has been supported  by the
Spanish  grants FPA2007-60275 and FPA2010-16963. R.B. has also been partially supported by the German Academic Exchange Service (DAAD).

\end{acknowledgments}

\appendix

\section{Phonon and kaon thermodynamical variables}
\label{freenergy}

The phonon and kaon thermodynamical variables needed for the computation of the bulk viscosities can be extracted from the
free energy of the CFL system, which reads
\begin{equation}
\label{freeE}
\Omega (\mu, \delta \mu_k, \delta \mu_{ph},T) = \Omega_q (\mu,T ) + \Omega_k (\mu, \delta \mu_k,T) + \Omega_{ph} (\mu, \delta \mu_{ph},T) \ ,
\end{equation}
where $\Omega_q$ is the quark contribution, while $\Omega_k$ and $\Omega_{ph}$ are the kaon and phonon contributions, respectively.
The free energy of the phonons and kaons in the presence of the ``fake" chemical potentials that characterize the out of equilibrium state is
\begin{equation}
\Omega_i (\mu,\delta \mu_i,T) = \frac{T}{2 \pi^2} \int^\infty _0 dp p^2 \ln{(1 - \exp(-(E_i - \delta \mu_i)/T))} \ , \qquad  i= k, ph \ .
\end{equation}
The $\mu$-dependence in the contribution to the free energy of the phonons and kaons appears because the energy of the quasiparticle depends on $\mu$.
In particular, for the phonons both the speed of sound $c_s$ and the parameter $B$ that enters in their dispersion relation, see Eq.~(\ref{newB}), depend on $\mu$.
For the kaons, both their velocity $v$,  mass $m_{K^0}$, and chemical potential $\mu_{K^0}^{eff}$ depend on $\mu$, although here we will only consider the dependence
of $\mu_{K^0}^{eff}$, as the  dependence of the other variables is unknown. 

The different particle densities are obtained from the free energy of the system
\begin{equation}
n_i = - \frac{\partial \Omega}{\partial \mu_i} \ , \qquad i =q, k, ph \ .
\end{equation}

From the free energy of the system we can also compute all the derivatives that appear in the expressions of the bulk viscosities. These are computed from
Eq.(\ref{freeE}), putting $\delta \mu_i= 0$ at the end of the computation. 

The phonon thermodynamical variables can be computed analytically. For the computation of the bulk viscosities that was done in Ref.~\cite{Mannarelli:2009ia}
it was realized that the phonon dispersion law at non-linear order was needed. The explicit expressions of the thermodynamical functions needed in our
computation can be found in that reference. The kaon thermodynamical variables have to be computed numerically.

It is interesting to note  that the following conditions are satisfied
\begin{equation}
\frac{\partial n_k}{\partial \mu} = \frac{\partial n_q}{\partial \mu_k} \ ,  \qquad \frac{\partial n_{ph}}{\partial \mu} = \frac{\partial n_q}{\partial \mu_{ph}} 
\ , \qquad \frac{\partial n_{ph}}{\partial \mu_k} = \frac{\partial n_k}{\partial \mu_{ph}} = 0 \ .
\end{equation}
Further,  we will use the relation \cite{Alford:2007rw}\begin{equation}
\frac{\partial n_k}{\partial \mu}  = - \frac{m_s^2}{2 \mu^2} \frac{\partial n_k}{\partial \mu_k}  \ .
\end{equation}

\section{Dimensions of the bulk viscosity coefficients}
\label{dimensions}

The bulk viscosity coefficients introduced by Gusakov in Son's formulation of the relativistic superfluid hydrodynamics are related to those
introduced by Khalatnikov by mass factors. More explicitly, one has
\begin{equation}
\zeta_1^{Kh} = \frac{\zeta^1}{m} \ , \qquad  \zeta_2^{Kh} = \zeta_2 \ , \qquad \zeta_3^{Kh} = \frac{\zeta_3}{m^2} \ .  
\end{equation}

If $M, L, {\cal T}$ refer to scales of mass, length and time, respectively, then the three bulk viscosity coefficients in
Gusakov's theory have the following dimensions
\begin{equation}
[\zeta_1] = M L^2 {\cal T}^{-1} \ , \qquad [\zeta_2] = M L^{-1} {\cal T}^{-1} \ , \qquad [\zeta_3] = M L^5 {\cal T}^{-1} \ . 
\end{equation}

In all the main part of the paper we worked using natural units. The plots are given in units of SI. For the conversion
of the values of the bulk viscosities from natural units to IS units it is useful to remember that
\begin{equation}
1\, {\rm MeV} = 1.78 \cdot10^{-27} {\rm g} \ , \qquad 1 \,{\rm MeV} = 1.97 \cdot10^{-11} {\rm cm} \ , \qquad  1\, {\rm MeV} = 6.58 \cdot 10^{-22} {\rm sec} \ .
\end{equation}

When the bulk coefficients are expressed in natural units they seem to behave almost similarly, that is, one can see that they
have a similar $T$ dependence. In the SI system, 
and due to the different scalings in length of the coefficients, the coefficients show a numerical behavior rather different,
 as shown in the plots of  Fig.~\ref{fig:zeta1_total} and Fig.~\ref{fig:zeta2_total}, the ultimate reason being the different scaling in $L$
 of every coefficient.

We note here that in the analysis of the r-modes of the CFL phase done in Ref.~\cite{Andersson:2010sh}, it was claimed that only
$\zeta_3$ is important for the computation of the damping of the r-modes. However, the dimensions of $\zeta_3$ in that manuscript have been
wrongly assumed to be the same as those of $\zeta_2$, which probably affects the final numerical results of the whole analysis when converting
the different magnitudes from natural units to SI units.

\end{document}